\def\spose#1{\hbox to 0pt{#1\hss}}

\def\multleft#1{\hbox to size{\vbox {\halign {\lft{##}\cr #1}}\hfill}\par}
\def\multright#1{\hbox to size{\vbox {\halign {\rt{##}\cr #1}}\hfill}\par}

\def\today{\ifcase\month\or January\or February\or March\or April\or May\or
      June\or July\or August\or September\or October\or November\or December\fi
      \space\number\day, \number\year}

\def\asec{$^{\prime\prime}$}

\newcommand{\Msolar}{\mbox{\,$\rm M_{\odot}$}}

\def\H2{\hbox{H$_{2}$}}
\def\Lya{Ly$\alpha$~}

\newcommand{\gtsim}{\mbox{{\raisebox{-0.4ex}{$\stackrel{>}{{\scriptstyle\sim}}
$}}}}
\newcommand{\ltsim}{\mbox{{\raisebox{-0.4ex}{$\stackrel{<}{{\scriptstyle\sim}}
$}}}}

\newcommand{\zp}{$z^{\prime}$}

\documentclass{mn2e}

\usepackage{times}
\usepackage{amssymb}
\usepackage{epsfig}

\begin{document}
\hsize=6truein
          
\title[Massive galaxies at redshifts $5<z<6$]
{The discovery of a significant sample of massive galaxies at redshifts
5$<$z$<$6 in the UKIDSS Ultra Deep Survey Early Data Release}

\author[R.J.~McLure et al.]
{R. J. McLure$^{1}$\thanks{Email: rjm@roe.ac.uk}, M. Cirasuolo$^{1}$, 
J. S. Dunlop$^{1}$, K. Sekiguchi$^{2}$, O. Almaini$^{3}$,
S. Foucaud$^{3}$,\and C. Simpson$^{4}$,  M.G. Watson$^{5}$, 
P. Hirst$^{6}$, M.J. Page$^{7}$, Ian Smail$^{8}$\\
\footnotesize\\
$^{1}$SUPA\thanks{Scottish Universities Physics Alliance} Institute
for Astronomy, University of Edinburgh, 
Royal Observatory, Edinburgh EH9 3HJ\\
$^{2}$Subaru Telescope, National Astronomical Observatory of Japan,
650 North A'ohoku Place, Hilo, Hawaii 96720, USA\\
$^{3}$School of Physics and Astronomy, University of Nottingham,
University Park, Nottingham NG7 2RD\\ 
$^{4}$Astrophysics Research Institute, Liverpool John Moores
University, Twelve Quays House, Egerton Wharf, Birkenhead CH41 1LD\\
$^{5}$Department of Physics \& Astronomy, University of Leicester, 
Leicester LE1 7RH\\
$^{6}$Joint Astronomy Centre, 660 N. A'ohoku Place, University Park,
Hilo, Hawaii 96720, USA\\
$^{7}$Mullard Space Science Laboratory, University College London, 
Holmbury St. Mary, Dorking, Surrey RH5 6NT\\
$^{8}$Institute for Computational Cosmology, Durham University, 
South Road, Durham DH1 3LE}

\maketitle

\begin{abstract}
We have exploited the large area
coverage of the combined UKIDSS Ultra Deep Survey (UDS) and Subaru/XMM-Newton
Deep Survey (SXDS) to search for bright Lyman-break galaxies (LBGs) at 
$z\geq5$. Using the available optical+near-infrared photometry to
efficiently exclude low-redshift contaminants, we identify nine
$z\geq5$ LBG candidates brighter than \zp$_{AB}=25$ within the 0.6
square~degree overlap region between the UDS early data release (EDR)
and the optical coverage of the SXDS. Accounting for selection
incompleteness, we estimate the corresponding surface density of 
$z\geq5$ LBGs with \zp$_{AB}\leq25$ to be $0.005\pm0.002$ per
square~arcmin. Modelling of the 
optical+near-infrared photometry constrains the candidates' redshifts
to lie in the range $5.1<z<5.9$, and provides estimates for their stellar masses. 
Although the stellar mass estimates are individually uncertain, a
stacking analysis suggests that the typical stellar mass of the LBG
candidates is $\gtsim\,5\times10^{10}\Msolar$ which, if confirmed,
places them amongst the most massive galaxies currently known at $z\geq5$.
It is found that $\Lambda$CDM structure formation can produce
sufficient numbers of dark matter halos at $z\geq5$ to accommodate our 
estimated number density of massive LBGs for plausible values of $\sigma_{8}$ and the ratio
of stellar to dark matter. Moreover, it is
found that recent galaxy formation models can also
account for the existence of such massive galaxies at
$z\geq5$. Finally, no evidence is found for the existence
of LBGs with stellar masses in excess of $\simeq3\times
10^{11}\Msolar$ at this epoch, despite the large co-moving volume surveyed.
\end{abstract}


\section{INTRODUCTION}

The increasing availability of survey fields with deep multi-wavelength
data has led to so-called ``dropout'' techniques becoming established
for photometrically identifying high-redshift galaxies. 
At redshifts of $z\simeq6$ it is now standard practice to select galaxy
candidates using the $i-$drop technique, a straightforward extension of
the Lyman-break method pioneered at $z\simeq3$ by Guhathakurta, Tyson
\&  Majewski (1990) and Steidel \& Hamilton (1992). Within this
context, the unique combination of depth 
and image quality provided by the Hubble Space Telescope (HST) has 
arguably made the largest contribution to our understanding of 
high-redshift galaxies. 
Analysis of the Hubble deep fields, the GOODS fields, and in
particular the Hubble ultra-deep field (HUDF), has led to the 
identification of hundreds of $i-$drop galaxy candidates at
faint ($z_{850}\geq26$\footnote{All magnitudes are quoted in the AB
system (Oke \& Gunn 1983)}) magnitudes (e.g. Bouwens et al. 2006;
Malhotra et al. 2005; Bunker et al. 2004; Dickinson et al. 2004). 
As a consequence, we now have a vastly improved understanding of the 
galaxies which undoubtedly dominate the star-formation density of the
Universe during the epoch immediately following reionisation.

However, although HST has greatly advanced our knowledge of
high-redshift galaxies, the various HST deep fields suffer from one
fundamental weakness. Crucially, their small area (e.g. the
HUDF covers $\simeq$13 square~arcmin) makes them subject to significant
cosmic variance (Somerville et al. 2004) and completely unsuitable for
investigating the most luminous, rarest objects.  Indeed, of the many
hundreds of $i-$drop galaxies identified in the deep HST fields, only
one has $z_{850}\leq25$ (SBM03$\#3$, spectroscopically
confirmed at $z=5.78$; Bunker et al. 2003). This is an important
shortcoming because accurately determining the number density of
massive galaxies at high-redshift ($z\geq5$) has the potential to
place important new constraints on galaxy formation models, only 1 Gyr
after the Big Bang. From an observational perspective, it is clear
that what is required is a combination of both depth and large area coverage.

In this respect, the study of $z\simeq6$ LBGs in the Subaru Deep Field
(SDF; Maihara et al. 2001) by Shimasaku et al. (2005) is perhaps the
most notable. Based on deep Subaru optical imaging covering 767
square~arcmin, Shimasaku et al. identified twelve bright LBGs with
\zp-band magnitudes in the range 25.4$<$\zp$<$26.6,  photometrically
selected to lie in the redshift range $5.6<z<6.2$. Moreover, by using
two medium-band filters ($z_{B}$ and $z_{R}$) which divide the Subaru
\zp-prime filter in half, Shimasaku et al. gained crucial information
on the UV slope of their high-redshift candidates. As a consequence,
Shimasaku et al. were able to select a very clean sample, which should
have negligible contamination from low-redshift interlopers. In addition, Ota et al. (2005)
performed a search for \zp$\leq26$ LBGs over a sky area of $\sim1$
square~degree in the SXDS field, using a primary selection criterion of
$i^{\prime}-z^{\prime}>1.5$. However, Ota et al. were only able to
strongly constrain the surface density of LBG candidates at
$z^{\prime}\geq25$, because at brighter magnitudes their lack of near-infrared 
data made it impossible to exclude significant contamination from
low-redshift interlopers, and cool galactic stars in particular.
Finally, it should be noted that there
is already evidence for a large population of $z\simeq6$
galaxies at fainter $z^{\prime}-$band magnitudes within the 
SXDS/UDS field. Using a combination of
the broad $i^{\prime}-$band SXDS imaging and a narrow band filter
($\lambda_c=8150$\AA), Ouchi et al. (2005) identified 515 potential
Ly$\alpha$ emitters (LAEs) within the redshift interval $5.65<z<5.75$, over an
area of one square~degree. Further deep optical
spectroscopy subsequently confirmed eight LAEs at $z\simeq5.7$ with 
asymmetric \Lya emission lines. However, the LAEs
detected by Ouchi et al. are fainter ($z^{\prime}\geq25$) than the objects
identified in this study (and presumably have lower mass).

In this paper we exploit two crucial advantages of the combined SXDS/UDS
data-set to explore a new area of LBG parameter space at
$z\geq5$. Firstly, the overlap region between the SXDS and the 
UDS EDR is $\geq2.5$ times larger than the SDF ($\simeq 0.6$
square~degrees), offering the prospect of
identifying the rarest $z\geq5$ LBGs with
$z^{\prime}\leq25$. Secondly, the near-infrared data from the UDS EDR 
allows us to effectively clean the sample of low-redshift
contaminants, and to obtain stellar mass estimates for the LBG candidates.
The structure of the paper is as follows. In Section 2 we give a
brief review of the optical and near-infrared data currently available
in the SXDS/UDS field. In Section 3 we outline the criteria
adopted for selecting the $5<z<6$ galaxy candidates, and discuss the
possible sources of contamination. In Section 4 we present the results of our modelling of the
candidates' spectral energy distributions (SEDs). 
In Section 5 we estimate the number densities of massive LBGs at
$z\geq5$ and compare with the predictions
of current galaxy formation models. In Section 6 we compare our
results with those in the recent literature, before presenting our 
conclusions in Section 7. Throughout the paper we adopt the following 
cosmology: $H_{0}=70$ km\,\,s$^{-1}$Mpc$^{-1}$, $\Omega_{m}=0.3$, 
$\Omega_{\Lambda}=0.7$. When referring to the $i$~and $z-$band
filters, $i^{\prime}$ and $z^{\prime}$ specifically refer to the
Subaru filters, whereas $i_{814}$ and $z_{850}$ are the equivalent 
HST filters.

\section{The data}
In this paper we identify potential massive galaxies at high redshift
by combining optical data from the Subaru telescope with near-infrared
data from the UK Infrared Telescope (UKIRT). The optical data were 
taken as part of the Subaru/XMM-Newton Deep Survey (Sekiguchi et al. 2005) while the
near-infrared data has been taken from the early-data release (Dye et
al. 2006) of the UKIRT Infrared Deep Sky Survey (Lawrence et al. 2006). The
properties of both data-sets are briefly outlined below.

\subsection{The Subaru/XMM-Newton Deep Survey}
The Subaru/XMM-Newton Deep survey (SXDS) primarily consists of optical and
X-ray imaging covering an area of $\simeq 1.3$~square degrees, centred
on RA=02:18:00, Dec=-05:00:00 (J2000). In addition, the SXDS
field has IRAC and MIPS data from the Spitzer SWIRE survey (Lonsdale
et al. 2003), 
deep 1.4 GHz radio observations from the VLA (Simpson et al. 2006a) and
850-micron sub-mm observations from SCUBA (Mortier et
al. 2005). Of primary concern
for this study is the deep optical imaging of the field undertaken
with Suprime-Cam (Miyazaki et al. 2002) on Subaru. The optical imaging
consists of 5 over-lapping Suprime-Cam pointings, and covers an area
of $\simeq 1.3$~square degrees. The whole field has been imaged in the
$BVRi^{\prime}z^{\prime}$ filters, to typical $5\sigma$ depths of
$B=27.5$, $V=26.7$, $R=27.0$, $i^{\prime}=26.8$ and $z^{\prime}=25.9$ 
(2\asec$-$diameter apertures). 

\subsection{The UKIDSS Ultra Deep Survey}
The Ultra Deep Survey (UDS) is one of five near-infrared surveys being
undertaken with WFCAM (Casali et al. 2006, in preparation) 
on UKIRT, which together comprise the UKIRT Infrared Deep Sky Survey (UKIDSS).
The UDS will take seven years to complete, and will 
provide ultra-deep $JHK$ imaging of a 0.8 square~degree area 
(centred on the SXDS field), to $5\sigma$ point-source limits of
$J=26.0$, $H=25.4$ and $K=25.0$. The UDS early-data release consists
of the first twelve hours of $JK$ imaging data and reaches 
$5\sigma$ limits of $J=22.5$ and $K=22.5$ within a 2\asec$-$diameter
aperture. For this paper we have used an improved version of stacking
and catalogue extraction constructed by the current authors, rather
than the default version in the WFCAM Science Archive. This will be 
described in Foucaud et al (2006). The authors of this paper are also 
members of the UKIDSS consortium, so we intend that our improved
stacking and source extraction will be made available to help to develop
the final WSA database. Due to a small change in
the UDS field centre shortly after the beginning of the survey, the 
current $JK$ imaging is not uniform over the entire 0.8 square~degree
field. Consequently, for the purposes of this paper, we restrict our
analysis to the $0.6$~square~degree central region which has uniform
$JK$ data and optical Subaru coverage.

\begin{table*}
\begin{center}

\caption{Names, optical positions and $2^{\prime\prime}-$diameter
aperture photometry for the final nine high-redshift LBG candidates (quoted limits are $1\sigma$). 
The consistent image quality of the optical+near-infrared data
(0.7\asec$<$FWHM$<$0.8\asec) means that differential aperture
corrections between bands are small ($\leq0.1$~magnitudes). All aperture magnitudes have
been confirmed manually to check for errors in the original catalogue
photometry. Four objects (MCD2, MCD5, MCD8 \& MCD9) have photometry calculated in a
$1.6^{\prime\prime}-$diameter aperture to avoid possible contamination
from nearby sources (as a result we note that formally MCD9 has a colour of
$R-z^{\prime}=2.90\pm0.27$, rather than $R-z^{\prime}\geq3.0$). Although the majority of the aperture
photometry in the $i^{\prime}$ and $z^{\prime}$ filters has
S/N $\geq10$, a minimum error of 0.1 magnitudes was adopted during the
SED fitting process to account for possible zero-point errors.}
\begin{tabular}{lccccccccc}
\hline
Source & RA(J2000)&DEC(J2000)&$B$ & $V$ & $R$ & $i^{\prime}$ & $z^{\prime}$ &$J$ &$K$\\
\hline
MCD1& 02:16:27.81 & $-$04:55:34.1   &$>29.25$ & $>28.40$ & $>28.70$ & $27.33\pm0.31$&$24.63\pm0.07$& $>24.28$&$24.22\pm0.73$ \\
MCD2& 02:16:57.29 & $-$04:52:00.8   &$>29.49$ & $>28.64$ & $28.92\pm0.74$ & $26.31\pm0.11$&$25.04\pm0.08$& $24.20\pm0.60$&$23.99\pm0.52$ \\
MCD3& 02:17:02.70 & $-$04:56:59.3   &$>29.25$ & $>28.40$ & $>28.70$ & $26.15\pm0.11$&$24.63\pm0.07$& $23.71\pm0.51$&$23.37\pm0.39$ \\
MCD4& 02:17:02.90 & $-$05:23:17.3   &$>29.25$ & $>28.40$ & $>28.70$ & $26.09\pm0.10$&$24.63\pm0.07$& $>24.28$&$23.68\pm0.50$ \\
MCD5& 02:18:26.56 & $-$05:01:51.0   &$>29.49$ & $>28.64$ & $>28.94$ & $26.35\pm0.11$&$25.00\pm0.07$& $24.15\pm0.58$&$23.53\pm0.37$ \\
MCD6& 02:18:47.30 & $-$05:21:04.9   &$>29.25$ & $>28.40$ & $28.38\pm0.60$ & $25.40\pm0.06$&$24.25\pm0.05$& $23.89\pm0.57$&$23.99\pm0.62$ \\
MCD7& 02:18:53.20 & $-$04:40:40.5   &$>29.25$ & $>28.40$ & $>28.70$& $25.96\pm0.10$&$24.64\pm0.07$& $24.28\pm0.75$&$23.17\pm0.34$ \\
MCD8& 02:19:00.46 & $-$04:44:58.7   &$>29.49$ & $>28.64$ & $27.54\pm0.26$ & $25.39\pm0.05$&$24.36\pm0.04$& $>24.52$&$>24.51$ \\
MCD9& 02:19:04.30 & $-$04:47:55.2   &$>29.49$ & $>28.64$ & $27.52\pm0.26$ & $25.67\pm0.06$&$24.62\pm0.05$& $23.92\pm0.49$&$23.91\pm0.49$ \\
\hline
\end{tabular}
\end{center}
\end{table*}

\section{High-redshift candidate selection}
Our sample of bright LBGs at $z\geq5$ was initially selected from 
the Subaru optical imaging, followed by SED fitting of the
Subaru+UDS data to isolate robust candidates. The optical selection
criteria were as follows: 
\begin{itemize}
\item{$z^{\prime}\leq25$}
\item{non-detection ($<2\sigma$) in the $B$ and $V-$bands}
\item{$R-z^{\prime}\geq3$}
\end{itemize}
\noindent
The magnitude cut at $z^{\prime}\leq25$ was chosen to fully exploit the
area of the UDS, and to isolate the most luminous/massive $z\geq5$
galaxies which cannot be studied using smaller-area surveys. 
Furthermore, from a practical point of view, for objects with
$z^{\prime}\geq25$ the shape of the SED long-ward of the
$z^{\prime}-$band cannot be properly constrained 
(with the early-release UDS data), making it impossible to identify
robust candidates (see Section 3.2). The requirements for non-detections in
the $B$ and $V-$bands, together with a red $R-z^{\prime}\geq3$ colour,
are designed to ensure our selection function has a sharp low-redshift
cut-off at $z=5$. At $z\geq5$ both the $B$ and $V-$bands sample
rest-frame wavelengths short-ward of Ly$\alpha$, and any genuine
high-redshift LBGs should not be detected at $>2\sigma$ significance
in these bands. Moreover, in the redshift interval $5<z<6$ the Subaru $R$ and
$z^{\prime}-$filters bracket the onset of absorption by the Ly$\alpha$
forest, and simulations demonstrate that a colour-cut of
$R-z^{\prime}\geq3$ effectively excludes galaxies at $z<5$. 
Finally, it should be noted that during
the optical selection process we did not actually impose the standard
$i-$drop criterion. The reason for this is that
the standard colour cut (i.e. $i_{814}-z_{850}>1.3$; Bunker et
al. 2004) can exclude galaxies in the redshift 
range $5.0<z<5.5$. In contrast, for the purposes of this study, we are 
interested in including all possible candidates at $z\geq5$, which is
better achieved using the $R-z^{\prime}\geq3$ colour-cut. However, as
can be seen from Table~1, all of the final nine LBG candidates
satisfy $i^{\prime}-z^{\prime}\geq1.0$, fully consistent with the expected
colours of $z\geq5$ LBGs in the Subaru filter set.

\begin{figure*}
\centerline{\epsfig{file=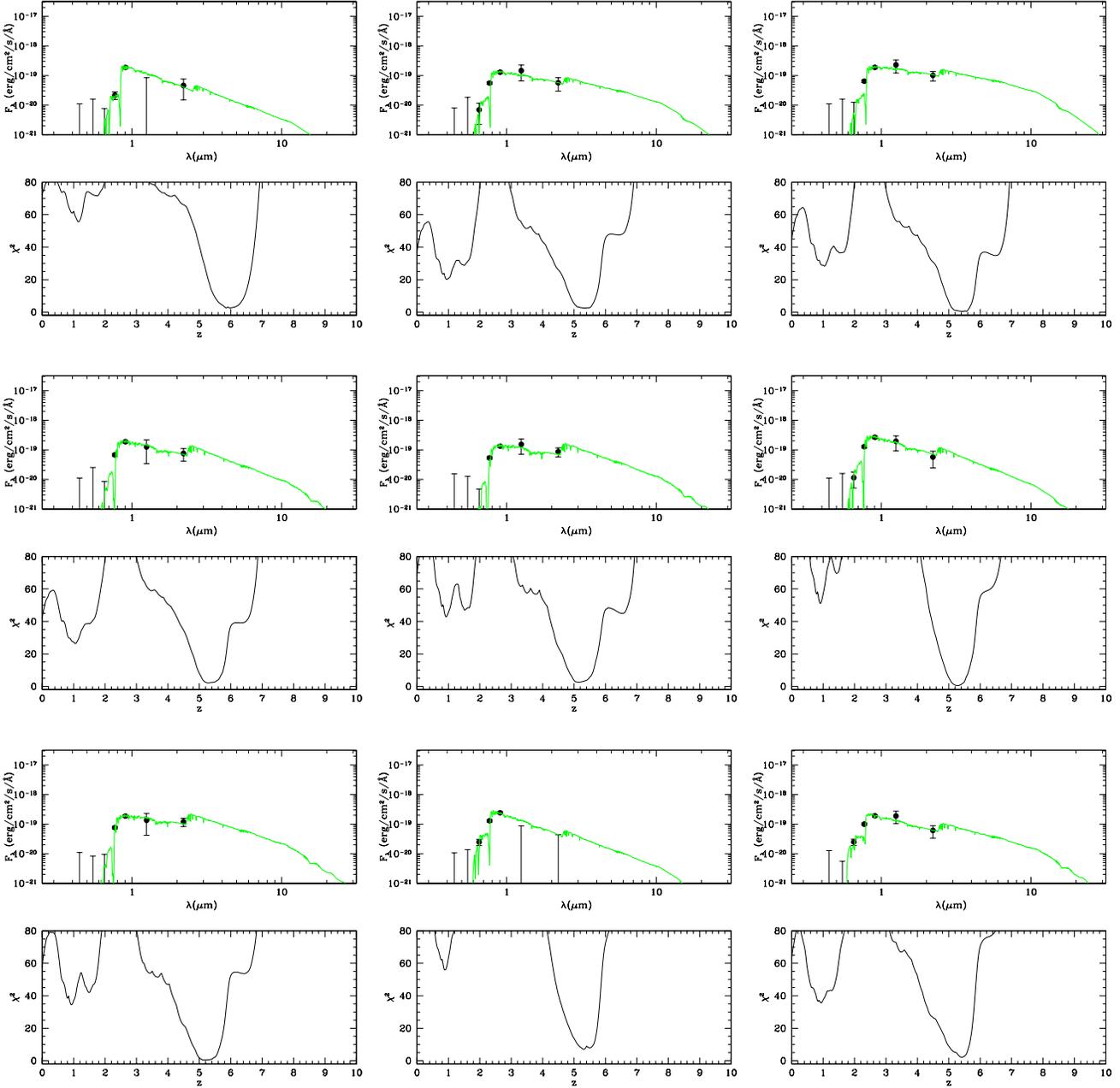,width=17.5cm,angle=0}}
\caption{The best-fitting SEDs to the photometry
of the final nine LBG candidates (see Table 1) in numerical order,
horizontally from the top left. The best-fitting parameters
are listed in Table 2. The lower panels show the value of $\chi^{2}$
for the fits as a function of photometric redshift (marginalised over
all other free parameters).} 
\end{figure*}

\begin{figure*}
\centerline{\epsfig{file=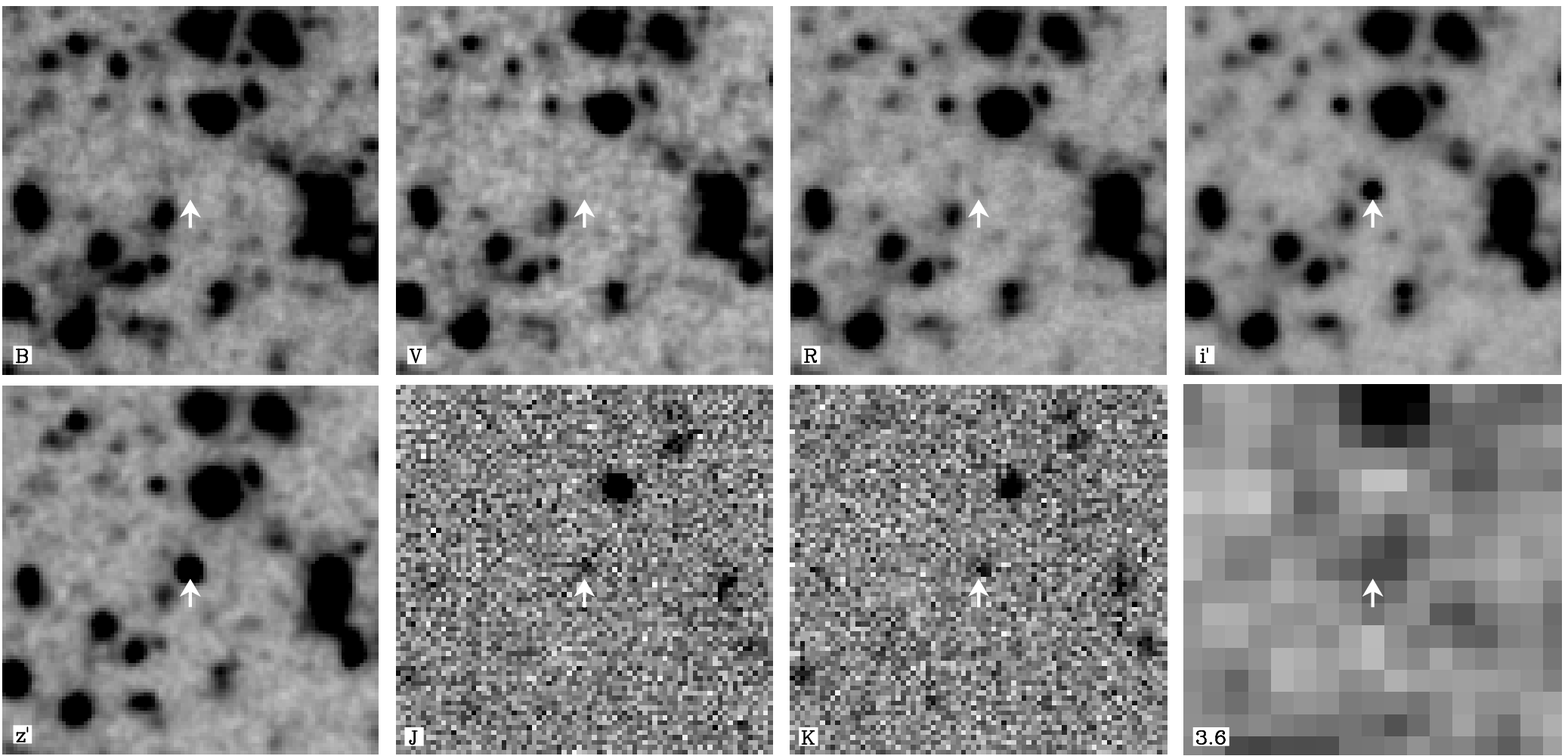,width=15.0cm,angle=0}}
\caption{{\it BVRi$^{\prime}z^{\prime}$JK} and $3.6\mu$m postage stamp
images of the stacked data for all nine of the final high-redshift LBG candidates. 
Each stamp is twenty arcsec square (except the $3.6\mu$m stamp which
is ten arcsec square), and is displayed using a grey-scale which is a linear 
stretch $\pm3\sigma$ around the mean sky level. The name of the filter
is shown in the bottom-left corner of each stamp, and the position of
the stacked LBG candidate is highlighted using an arrow.}
\end{figure*}

\subsection{Optical+near-infrared model fitting}
Following the application of the initial optical selection criteria,
the next step in the selection process was to analyse each
potential candidate with our template-fitting photometric redshift
code (Cirasuolo et al. 2006, in preparation), which is largely based on the public package {\sc hyperz} 
(Bolzonella, Miralles \& Pell\'{o} 2000). The photometric redshift for
each galaxy is computed by fitting the observed multi-wavelength ($
BVRi'z'JK$) photometry with synthetic galaxy templates. We adopt the
stellar population synthesis models of Bruzual \& Charlot (2003) for
the galaxy templates, assuming a Salpeter initial mass function (IMF) with a lower and upper
mass cutoff of 0.1 and  100 $M_{\odot}$ respectively. Both instantaneous burst and exponentially
declining star formation models -- with e-folding times in the range $\rm 0.1 \leq \tau (Gyr)\leq 15$ --
were included, all with fixed solar metallicity. The code accounts for
dust reddening by following the Calzetti et al. (2000) obscuration law
within the range $0 \leq A_V \le 10$, and accounts for Lyman series
absorption due to the HI clouds in the inter galactic medium according
to the Madau (1995) prescription. In addition to the best-fitting
photometric redshift, the code also returns best-fit values for: SED type,
age, mass and reddening. 

In addition to providing redshift and stellar mass estimates, the
principle reason for performing the optical+near-infrared SED fits 
was to clean the sample of low-redshift interlopers. A simple 
application of the optical selection criteria described above produced
an initial sample of 74 candidates. By making full use of the extra
information provided by the near-infrared UDS data, 65 members of this
initial sample were eventually excluded as low-redshift contaminants.
The names, positions and photometry for the final nine LBG candidates are listed
in Table 1, and the fits to their optical+near-infrared photometry 
are shown in Fig 1. In the next section we discuss in detail the
potential sources of contamination, and the strategies adopted for 
excluding them from the final sample.


\subsection{Sources of contamination}
In the current study there are three types of contaminants which are
of potential concern: ultra-cool galactic stars (M, L or T dwarfs),
extremely red objects (EROs) at $z\simeq1$ and high-redshift quasars. 
It is important to develop strategies for excluding these contaminants based on their
optical+near-infrared SEDs because, unlike previous HST studies,
ground-based resolution makes it impossible to exclude stellar and QSO
contaminants on the basis of being spatially unresolved.

\subsubsection{Ultra-cool galactic stars}
Due to the $\simeq0.8$\asec\, FWHM resolution of the available
ground-based imaging of the SXDS/UDS field, high-redshift LBGs are
expected to be unresolved point sources. Consequently, the prospect of
significant contamination of the final sample by ultra-cool galactic stars is 
potentially serious. Ultra-cool galactic stars (M, L or T dwarfs)
perfectly mimic the redshifted SEDs of $z\geq5$ LBGs in the optical (Hewett et al. 2006),
and would certainly pass our initial optical selection
criteria. Consequently, in order to exclude stellar
contaminants it is necessary to rely on the fact that ultra-cool
galactic stars exhibit redder $z^{\prime}-J$ colours than genuine
$z\geq5$ LBGs. To quantify this we used the library of
spectra maintained by Sandy Leggett
\footnote{www.jach.hawaii.edu/skl} to 
calculate the likely 
colours of M, L and T dwarfs in the Suprime-Cam+WFCAM filter set.
This process revealed that we could efficiently exclude the cooler L
and T dwarfs from the final sample by rejecting any potential
candidate which was detected in the UDS EDR $J-$band imaging at
$\geq2\sigma$ significance. This is because the red
$z^{\prime}-J\,\gtsim\,1.5$ colours typical of L and T dwarfs means 
that, at $z^{\prime}\leq25$, they should be significant ($\geq3\sigma$)
detections in the EDR $J-$band imaging. However, unfortunately the
potential for contamination by M dwarfs is much more significant.
 Due to the fact that none of the final nine LBG candidates are
formally detected in the $J-$band (all $<2\sigma$ detections), the
limits on their $z^{\prime}-J$ colours are typically
$z^{\prime}-J\,\ltsim\,1.0$, compatible with those of M dwarfs with 
$i^{\prime}-z^{\prime}\geq1.0$. To tackle this issue, as part of the 
SED fitting we used the available M dwarf spectra
(spectral types M1 to M9) as templates to fit the LBG candidates' 
photometry. For seven of the nine final candidates, using galaxy
templates produced significantly better fits to the LBG photometry (Table 2) than
could be achieved with the M dwarf templates. However, for two of the
final candidates (MCD2 \& MCD3) the M dwarf SED fits were of
comparable quality to those achieved with $z\geq5$ galaxy
templates. Consequently, it remains possible that these two objects
are M dwarf contaminants. However, we have chosen to keep them in the
final sample because $z\geq5$ LBG templates provide a very good fit
to their photometry and, even including these two objects, a stacking
analysis suggests the final candidate list of nine objects is not
heavily contaminated by M dwarf stars.

\begin{figure*}
\centerline{\epsfig{file=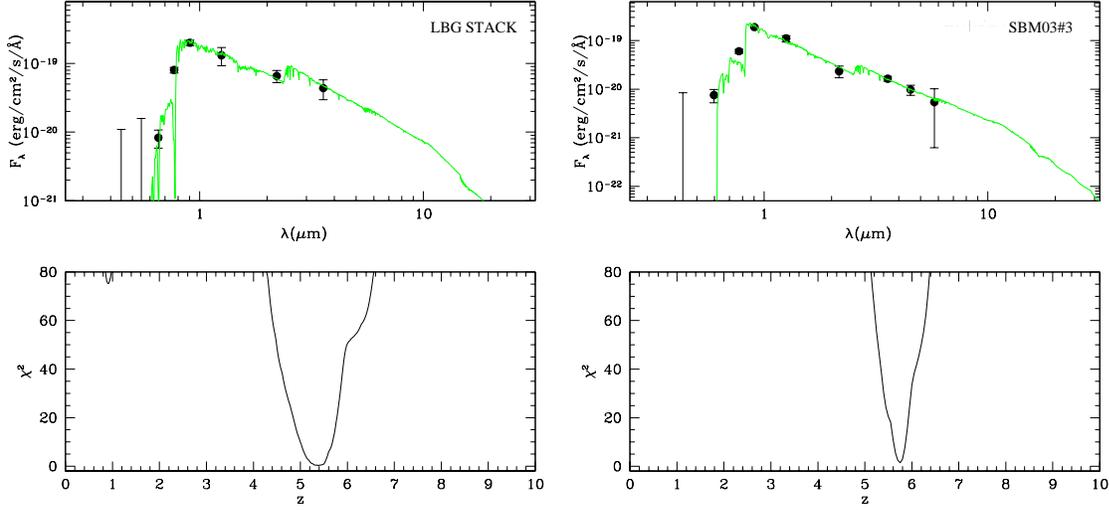,width=15.0cm,angle=0}}
\caption{The left-hand panel shows the best-fitting SED to the stacked
photometry of the final nine high-redshift LBG
candidates. Due to the extra depth provided by stacking, the SED of
the LBG stack includes a detection at $3.6\mu$m from the SWIRE survey
data. Furthermore, the LBG stack remains undetected in the $B$ and
$V-$bands to $1\sigma$ limits of $B=30.25$ and $V=29.45$. As discussed
in the text, the SED of the LBG stack is not compatible with that of a
$z\simeq1$ ERO or an ultra-cool galactic star, strongly
suggesting that the final LBG candidate list is not heavily
contaminated. The right-hand panel shows
our SED fit to the brightest LBG known in the GOODS CDFS (SMB03$\#03$)
which has a spectroscopic redshift of $z=5.78$ (Bunker et al.~2003),
where the photometry has been taken from the GOODS MUSIC catalogue 
(Grazian et al. 2006). Our SED fit recovers the
spectroscopic redshift $z_{phot}=5.7\pm0.1$, and has an age and stellar mass
($5\times10^{10}\Msolar$) very similar to the SED fit performed by
Eyles et al. (2005). The similarity between the SEDs of the LBG stack and
SMB03$\#03$ provides further evidence of the reliability of our
selection technique.}
\end{figure*}

\subsubsection{Stacking analysis}
In order to confirm that the final sample of nine LBG candidates is
not heavily contaminated by galactic M-dwarf stars, we stacked (averaged)
the images of the final nine candidates in all available
bands. Postage stamp images centred on the stack of all
nine high-redshift LBG candidates in the $BVRi^{\prime}z^{\prime}JK$
and 3.6$\mu$m Spitzer bands are shown in Fig 2, and the SED fit to the 
stacked photometry is shown in Fig 3. 

Several points regarding the stacked image and accompanying 
SED fit are noteworthy. Firstly, due to the extra $\gtsim 1$
magnitude in depth achieved by stacking, it is possible to
confirm that the average SED of the final candidates is robustly 
undetected in the bluest optical bands to $2\sigma$ limits of $B=29.5$
and $V=28.7$. Secondly, the colours of the LBG stack are 
$i^{\prime}-z^{\prime}=1.28\pm0.14$ and $z^{\prime}-J=0.22\pm0.35$. An M-dwarf star with
$i^{\prime}-z^{\prime}=1.3$ is expected to have  $z^{\prime}-J\simeq
1.0$, significantly redder than the $z^{\prime}-J$ colour of the LBG stack. 
This alone is enough to strongly suggest that the final candidate list
is not heavily contaminated by M-dwarfs, and is confirmed by the fact
that the best-fit to the stacked $BVRi^{\prime}z^{\prime}JK$
photometry using an M-dwarf template can be confidently excluded ($\Delta \chi^{2}=25$).
Moreover, although each of the
final nine candidates is individually undetected in the 3.6$\mu$m
SWIRE data, the stack does provide a marginal $\simeq1.5\sigma$ detection
($m_{3.6}=24.0\pm0.5$), entirely consistent with a prediction based
on the individual SED fits shown in Fig 1. Crucially, if the final
candidate list consisted largely of ultra-cool galactic stars, there
would be no detection at 3.6$\mu$m. 

There is one final comparison which can be performed using the stacked
data to provide further information on the reliability of our
selection technique. As previously mentioned, in the GOODS CDFS field
there is one bright $z\geq5$ LBG known with $z_{850}\leq25$ (SBM03$\#03$),
spectroscopically confirmed to be at $z=5.78$ by Bunker et
al. (2003). This object is the only high-redshift LBG within the deep
HST survey fields which is comparable in brightness to the candidates
selected here. In the right-hand panel of Fig~3 we show the photometry
for this object, together with our SED fit. Reassuringly, our SED fit
accurately recovers the spectroscopic redshift
($z_{phot}=5.7\pm0.1$) using best-fitting model parameters very
similar to those required  to fit the photometry of our
nine high-redshift LBG candidates. Moreover, it can clearly be seen
from Fig~3 that the SEDs of SBM03$\#03$ and our LBG stack are very
similar, providing additional evidence that our final candidate list is robust. 

\subsubsection{Extremely red objects}
Based solely on our optical selection criteria, contamination of the
final sample by EROs at $z\simeq1$ is possible if the 4000\AA\, break
is confused with the Lyman break. However, given the extra information
from the UDS EDR near-infrared data, we can be confident that ERO
contamination is not an issue. As can be seen from Fig~1, none of the 
final nine candidates displays a problematic degeneracy between plausible 
solutions at low and high redshift. Although an alternative
photometric redshift solution at $z\simeq1$ inevitably exists for each 
candidate, in every case the low-redshift solution can be excluded 
at $\gg99.9$\% confidence ($\Delta \chi^{2}\geq18$). This is confirmed
by our SED fit to the stacked LBG photometry (Fig 3) which shows the
ERO solution at $z\simeq1$ to be completely unacceptable ($\Delta \chi^{2}=75$).
This clear separation between
high and low-redshift solutions is particularly notable given that
during the SED fitting process, extreme values of intrinsic reddening
($0<A_{V}<10$) were fully explored.

From a more empirical perspective, we have also investigated the
potential for contamination using a sample of 3715 $R-K\geq~5.3$ (Vega)
EROs with a $K-$band limit of $K_{tot}\leq22.2$, selected from the
same data-set as our LBG candidates (Simpson et al. 2006b). Within this
sample there are 2140 EROs with photometric redshifts in the range
$0.8<z<1.4$ and $z^{\prime}\leq25$. As expected, none of these EROs
display a statistically acceptable photometric redshift solution at $z\geq5$. 
The fundamental reason for this is that none of them have
the blue SED slope long-ward of the $z^{\prime}-$band displayed by our
final $z\geq5$ LBG candidates. Perhaps the most convincing evidence
for this comes from the SED of the stacked photometry of all nine
$z\geq5$ LBG candidates (Figs 2 \& 3), which has a colour of 
$z^{\prime}-K=0.73\pm0.25$. For comparison, the average
$z^{\prime}-K$ colour of $z\simeq1$ EROs in the UDS field is
$z^{\prime}-K=2.32\pm0.32$ (where the quoted uncertainty is the 
standard deviation).

\subsubsection{High-redshift quasars}
High-redshift quasars in the redshift range $5.0<z<6.5$ have predicted
$i^{\prime}-z^{\prime}$ and $z^{\prime}-J$ colours which are
comparable with those of our final nine LBG candidates (Willott et
al. 2005). However, none of the final candidates are detected in the deep $0.2-12$~keV band
X-ray imaging of the SXDS with XMM-Newton. The sensitivity of the
XMM-Newton SXDS observations varies somewhat over the SXDS field, but the
typical X-ray flux upper limits imply X-ray luminosity limits of $\simeq
(5-10)\times10^{44}$ erg s$^{-1}$ (rest-frame energy band $\sim1-70$
keV). Moreover, based on the latest constraints on the QSO
luminosity function at $z\simeq6$ (Willott et al. 2005), contamination
of our sample from high-redshift QSOs is expected to be minimal 
(conservatively $\leq1$ QSO is anticipated with $z^{\prime}\leq25$
over the SXDS/UDS field). 

\subsubsection{Lensing}
Finally, given that we have deliberately selected the brightest objects of
their type at $z\geq5$, it is important to at least consider the
possibility that some of the final LBG candidates may be lensed. 
The first scenario to consider is the prospect of moderate
amplification due to a lensing geometry where the LBG candidates lie 
outside the Einstein radii of lensing objects along the
line-of-sight. If we model the potential lens as a single isothermal
sphere, then the amplification is :
$\mu=\frac{\theta}{\theta-\theta_{E}}$, where $\theta_{E}$ is the
Einstein radius $\theta_{E}=4\pi(\frac{\sigma_{V}}{c})^2\frac{D_{LS}}{D_{S}}$,
$\sigma_{V}$ is the velocity dispersion of the lensing galaxy, $D_{S}$
is the angular diameter distance of the source and $D_{LS}$ is the
angular diameter distance of the source from the lens. Four of the final nine
LBG candidates have nearby companions within $2.5$\asec of the
line-of-sight, which could potentially produce amplification via
lensing. To investigate this we have estimated the velocity dispersions and
consequent amplification for these objects, using their photometric
redshifts and the $M_{i}-\sigma_{V}$ correlation at $z\simeq0$ from 
Bernardi et al. (2003). This calculation reveals that none of the
companion objects are compatible with having $\sigma_{V}\geq100$~km~s$^{-1}$, and that any
amplification from lensing is therefore small (i.e.~$\mu\leq1.1$).

The second scenario to consider is the prospect of high amplification 
due to extremely close alignment along the line-of-sight between the
LBG candidates and potential lensing objects in the
foreground. Although difficult to rule-out completely, the remarkably
clean nature of the LBG candidates' optical photometry suggests that it
is quite unlikely. For the $z\geq5$ LBG candidates the largest cross-section for
lensing occurs at $z\simeq0.8$. Consequently, if we consider an
$L^{\star}$ elliptical at $z=0.8$ as the potential lens, we can
predict that it would have apparent magnitudes of $K\simeq19.9$,
$V\simeq25.5$ and $B\simeq26.6$ (assuming $M^{\star}_{K}=-22.9$; 
Pozzetti et al. 2003). These magnitudes are completely incompatible
with those of the stack of the final nine LBG candidates, which is 
undetected to $2\sigma$ limits of $B=29.5$ and $V=28.7$, and
has a $K-$band magnitude of $K=23.86\pm0.23$. Consequently, the average
luminosity of any potential lensing objects would have to be
$\ltsim\,0.03\,L^{\star}$ to avoid detection in the stacked photometry,
which leads us to conclude that the probability of lensing is negligible.

\subsubsection{Colour-colour selection}
Finally, for completeness, in Fig 4  we show the location of the final nine
high-redshift LBG candidates on the $R-z^{\prime}$ versus
$z^{\prime}-J$ colour-colour plane. Also shown in Fig 4 is the locus
of M, L and T dwarfs and those EROs from Simpson et al. (2006b) which
satisfy our selection criteria of being undetected in the $B$ and
$V-$bands, and have $z^{\prime}-$band  magnitudes in the same range as the
LBG candidates ($24<z^{\prime}<25$). Although Fig 4 suggests that 
traditional colour-colour selection based on $R-z^{\prime}\geq3$ and
$z^{\prime}-J\leq1$ would successful isolate our high-redshift LBG
candidates from low-redshift EROs, it also illustrates the potential
for contamination by M dwarf stars (as discussed in Section
3.2.1). However, once again, the location of the stacked LBG
photometry on the $R-z^{\prime}$ versus $z^{\prime}-J$ plane strongly
suggests that the final LBG candidate list is not significantly contaminated
by low-redshift interlopers. For the purposes of obtaining a robust
sample of high-redshift LBG candidates, Fig 4 further emphasises the
importance of modelling the full $BVRi^{\prime}z^{\prime}JK$ SEDs,
using both galaxy and stellar templates, rather than relying on simple 
colour-colour selection.

\begin{figure}
\centerline{\epsfig{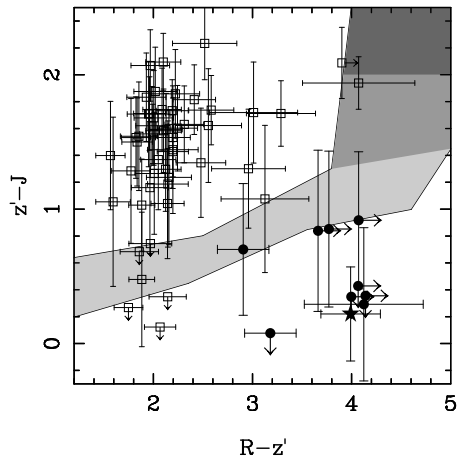}}
\caption{The $R-z^{\prime}$ versus $z^{\prime}-J$ colour-colour
plane. The final nine high-redshift LBG candidates are shown as filled
circles (small off-sets have been applied to several of the candidates
to prevent overlapping). The filled star shows the location of the
stacked LBG photometry. The open squares are $R-K>5.3$ (Vega) EROs
selected from the UDS EDR (Simpson et al. 2006b) which are undetected
in the $B$ and $V-$bands and have $z^{\prime}-$band magnitudes in the
same range as the LBG candidates ($24<z^{\prime}<25$).
The shaded area shows the locus of M, L and T dwarfs
(light, medium and dark grey shading respectively) based on the data of Hawley et
al. (2002) and West, Walkowicz \& Hawley (2005).}
\end{figure}

\begin{table*}
\begin{center}
\caption{The best-fitting parameters from the optical+near-infrared SED fitting. 
Columns list the name, SED template, $\chi^{2}$, primary redshift solution, $1\sigma$ redshift range, extinction, age, 
stellar mass and allowed stellar mass range (see text for
discussion). Although a wide range of different starformation
histories were considered, in all cases the best-fitting SED template
was either 1=instantaneous burst or 2=exponential decay model (e-folding time $\tau=0.3$ Gyr). The
photometry for the two objects highlighted with a $^{\star}$ symbol
can be fitted with an M dwarf stellar template with a comparable $\chi^{2}$, and these objects cannot be firmly ruled out as possible stellar contaminants.}
\begin{tabular}{lcccccccc}
\hline
Source &SED&$\chi^{2}$&z& $\Delta z$ &A$_{V}$&Age/Myr &
Log(${\rm M}/\Msolar)$& $\Delta$M\\
\hline

MCD1 &1\phantom{0}&2.6&5.85& 5.76$<z<$6.20&0.0&\phantom{0}57& 10.3 & $10.3<{\rm M}<10.8$\\

MCD2 &1$^{\star}$&2.5&5.38& 5.08$<z<$ 5.56&0.8&\phantom{0}57&10.8  & $10.3<{\rm M}<10.9$\\

MCD3 &1$^{\star}$&0.4&5.47& 5.16$<z<$ 5.62&1.0&\phantom{0}57&11.2  & $10.6<{\rm M}<11.3$\\

MCD4 &1\phantom{0}&2.1&5.29& 5.16$<z<$ 5.54&0.0&181&11.0 &$10.5<{\rm M}<11.1$\\

MCD5 &1\phantom{0}&2.5&5.13& 5.00$<z<$ 5.38&0.0&255&11.1 &$10.6<{\rm M}<11.3$\\

MCD6 &1\phantom{0}&0.5&5.26& 5.12$<z<$ 5.42&0.0&114&10.6 &$10.3<{\rm M}<10.7$\\

MCD7 &1\phantom{0}&0.4&5.18& 5.02$<z<$ 5.44&0.4&181&11.3 &$10.7<{\rm M}<11.4$\\

MCD8 &1\phantom{0}&7.0&5.31& 5.22$<z<$ 5.38&0.0&\phantom{0}57&10.0&$\phantom{0}9.9<{\rm M}<10.1$\\

MCD9 &2\phantom{0}&2.1&5.41& 5.30$<z<$ 5.54&0.6&509&11.2&$10.6<{\rm M}<11.5$\\

\hline
\end{tabular}
\end{center}
\end{table*}

\section{modelling results}
In Table 2 we list the best-fitting galaxy template parameters
resulting from the optical+near-infrared SED fits to each of the final nine
high-redshift LBG candidates (as illustrated in Fig 1). All of the
candidates have best-fitting photometric redshifts in the range
$5.1<z<5.9$, validating our original selection criteria. 
Moreover, as previously discussed, none of the candidates displays a
plausible low-redshift solution, despite the fact that an extreme
range of intrinsic reddening ($0<A_{V}<10$) was explored during the
SED fitting process. This is potentially
crucial because, as highlighted by Dunlop, Cirasuolo \& McLure (2006),
it is possible to confuse extremely dusty ($A_V\simeq4-6$) low-redshift
objects with genuine high-redshift galaxies if the range of reddening
explored is constrained to moderate values (e.g. $A_V\leq2$).

\subsection{Ages and reddening}
Although the age of the Universe at $z\simeq5$ ($1.1$ Gyrs) was not used as a prior in
the model fitting, reassuringly the best-fitting ages for all the
final nine candidates naturally satisfy this constraint, and are
consistent with formation redshifts in the range $5.6<z_{for}<9.0$. As is
expected given the LBG selection criteria, there is no evidence in
our sample for substantial amounts of reddening. Indeed, the range of
reddening displayed by our sample
($0<A_V<1$) is perfectly consistent with that found for LBGs at
$z\simeq3$ by Shapley et al. (2001). 

\subsection{Stellar masses}
By concentrating exclusively on the brightest ($z^{\prime}\leq25$)
candidates, this study was designed to investigate the most massive LBGs at $z\geq5$. 
The success of this policy can be seen from Table 2, which shows that
six of the nine final candidates have estimated stellar masses $\gtsim\,
5\times 10^{10}\Msolar$, and five have estimated masses $\gtsim\, 10^{11}\Msolar$. 
If these stellar mass estimates are accurate, then the LBG candidates
are among the most massive galaxies yet discovered at these redshifts 
(see discussion in Section 5), and are comparable to the highest
stellar masses found for LBGs at $z\simeq3$ by Shapley et al. (2001), 
who also adopted a Salpeter IMF. In fact, these stellar mass estimates
suggest that several of the $z\geq5$ LBG candidates
have already built-up a stellar mass comparable with 
M$_{\rm{stars}}^{\star}$ today ($\simeq10^{11}\Msolar$; Cole et al. 2001). 

However, given the current depth of the near-infrared data from the
UDS EDR, and the fact that the SWIRE data covering the UDS field is
not deep enough to individually detect the LBG candidates, it is clear
that on an object-by-object basis the stellar mass estimates must be
regarded with some caution. The first issue to consider is the
range of stellar masses which are allowable within the SED templates
we have adopted to fit the optical+near-infrared photometry. By
identifying the SED templates with the lowest and highest masses which 
still provide a statistically acceptable fit to the observed photometry 
(conservatively $\Delta \chi^{2}\leq10$, marginalised over all other
free parameters), we have calculated the allowable stellar mass range
for each of the LBG candidates (Table 2). This calculation reveals
that the photometry of most of the LBG candidates can be acceptably 
reproduced (although with a worse $\chi^{2}$) by SED templates with a 
factor of $\simeq 3$ less stellar mass. 

In addition, there is also the
added uncertainty introduced by the choice of a specific IMF. For the SED fits
presented in this paper we have made the standard choice of adopting a
Salpeter IMF, for ease of comparison with previous results in the
literature. However, it is widely recognised that the Salpeter IMF
results in higher stellar mass estimates than other popular
choices. For example, re-fitting the LBG optical+near-infrared photometry
with a Kennicutt or Chabrier IMF  would produce stellar mass
estimates a factor of $\simeq1.5$ lower. In conclusion, it is clear that
the individual stellar mass estimates for the LBG candidates are very
likely uncertain to within a factor of $\simeq5$.

In order to obtain a more robust estimate of the typical mass of
the $z\geq5$ LBGs, we have also calculated stellar mass
estimates based on the stacked LBG photometry. The obvious advantage
of this is that the extra depth provided by stacking produces much
more robust photometry in the $J$ and $K-$bands, and
also provides a detection at 3.6$\mu$m from the stacked SWIRE data. 
Using a straightforward average stack of the LBG data (see Figs 2 \&
3) our SED fit returns a stellar mass estimate of
$\simeq5\times10^{10}\Msolar$. However, this stack is biased due to the
inclusion of the two LBG candidates with the lowest stellar mass
estimates (MCD1 \& MCD8), which have the faintest $J+K$ photometry. 
This is confirmed by an SED fit to a median stack of the LBG photometry, which returns 
a stellar mass estimate of $\simeq1\times10^{11}\Msolar$ (as anticipated from the results in Table
2). In conclusion, taking into account systematic differences due to
the choice of IMF, the evidence from stacking the LBG data suggests a
typical stellar mass of $\gtsim\,5\times10^{10}\Msolar$.

\section{massive galaxies at high redshift }
The results presented in the previous section suggest that, despite
the considerable uncertainties, the typical stellar mass of the
$z\geq5$ LBG candidates is $\gtsim\,5\times10^{10}\Msolar$. In this section we
investigate whether the existence of such massive galaxies at
this early epoch is consistent with $\Lambda$CDM structure formation 
and current galaxy formation models. Within this section we adopt the
best-fitting stellar mass estimates for each LBG candidate, based on a
Salpeter IMF. However, when necessary, we adjust our estimated number
densities to account for differences due to specific choices of IMF.

\subsection{Surface density}
The first, model independent, quantity of interest to calculate is the surface 
density of $z^{\prime}\leq25$ galaxies at $z\geq5$. However, in order to
accurately compute the surface density, it is first necessary to
account for the inevitable incompleteness introduced during the
construction of the original SExtractor (Bertin \& Arnouts 1996) catalogues due to object
blending. To quantify this effect we performed simulations based on introducing
1000 fake LBG candidates at a time into the $z^{\prime}-$band images,
and attempting to recover them using the same SExtractor
configuration adopted for constructing the original $z^{\prime}-$band
catalogues. This immediately revealed that on average we lose
$\simeq10\%$ of possible LBG candidates due to object blending on the
$z^{\prime}-$band images. In addition, we then proceeded to run
SExtractor in two-image mode on the corresponding $B$ and $V-$band
data, using the $z^{\prime}-$band as the detection image. 
This revealed that we lose further LBG candidates due to
contamination of the matched apertures in the $B$ and $V-$bands by nearby
companions. In around $\simeq5\%$ of cases this contamination is
sufficient to fail our selection criteria requiring $\leq2\sigma$
detections in both the $B$ and $V-$bands. 

Consequently, when calculating the surface density
of $z^{\prime}\leq25$ LBGs it is necessary to scale-up by a factor of
$\simeq1.2$. Allowing for this correction factor, our estimate for
the surface density of $z^{\prime}\leq25$ LBGs at $z\geq5$ is 
$0.005\pm0.002$ per square~arcmin.  We note that this figure is
compatible with, although obviously much more robust than, a prediction based on the
one previously known $z\geq5$ LBG with $z_{850}\leq25$ in a compilation
of deep HST survey fields; $0.004\pm0.004$ per square~arcmin (Bouwens
et al. 2006). We also note that our final list of nine $z^{\prime}\leq25$ LBGs at
$z\geq5$ is fully consistent with the previous findings of Shimasaku
et al. (2005) in the SDF. Within the 767 square arcmin area of the
SDF, Shimasaku et al. did not find any possible $z^{\prime}\leq25$ LBGs, but did
identify 12 candidates in the magnitude range $25.4<z^{\prime}\leq26.6$. However,
it is important to remember that the selection technique adopted by 
Shimasaku et al. was tuned to identify LBGs in the redshift interval 
$5.6<z<6.2$. Only one of our final list of nine candidates lies within 
this redshift range (MCD1). Consequently, based on our final sample, we would predict
only $\simeq0.4$ LBG candidates with $z^{\prime}\leq25$ within the SDF
area which would satisfy the Shimasaku et al. selection criteria,
entirely consistent with their finding of none.

\subsection{Number densities}
The best-fitting stellar mass estimates listed in Table 2 suggest that
five of the LBG candidates have masses of M $\gtsim\,10^{11}\Msolar$.
In order to convert the estimated surface density of M
$\gtsim\,10^{11}\Msolar$ LBGs at $z\geq5$ into a number density, it is
necessary to compute the effective co-moving volume of our survey
(defined by the survey area and the redshift intervals within which 
each of the LBG candidates could have been detected; $z_{min}<z<z_{max}$). Due to our
adopted optical selection criteria, our selection function
has a sharp low-redshift cut-off at $z_{min}=5$. To compute $z_{max}$ we have
taken the SED fit to each candidate and redshifted it (keeping age,
reddening and mass constant) until the candidate fades below our 
magnitude limit ($z^{\prime}=25$). This calculation reveals that, for the five LBG
candidates with M $\gtsim\,10^{11}\Msolar$, the effective volume of the
survey is equivalent to the full co-moving volume covered by our 0.6
square~degree area over the redshift interval $5.0<z<5.6$
(i.e.~$3.3\times10^{6}$Mpc$^{3}$).

\begin{figure}
\centerline{\epsfig{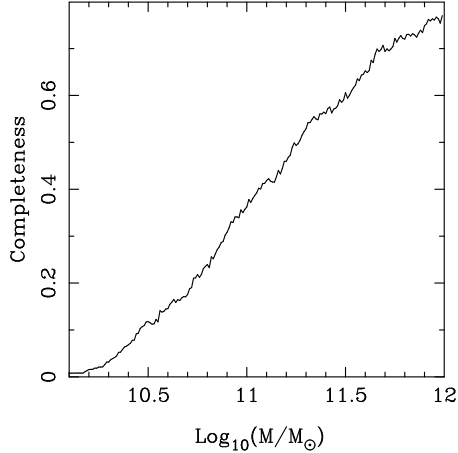}}
\caption{The estimated completeness of our LBG selection criteria as a
function of stellar mass, based on subjecting a catalogue of simulated LBG photometry
to the same selection criteria applied to the real data. It can be seen
that at stellar masses of $\simeq10^{11}\Msolar$ our sample is
expected to be $\simeq40\%$ complete.}
\end{figure}

\subsubsection{Completeness}
Additionally, in order to calculate a meaningful estimate of the
number density of M $\gtsim\,10^{11}\Msolar$ LBGs in our sample, it is also
necessary to estimate our completeness as a function of stellar mass. 
To investigate this we produced simulated photometry for 10,000 LBGs with
parameters randomly selected from uniform distributions with the following ranges:
\begin{itemize}
\item{Stellar masses: $10^{10}\Msolar<$ M $<10^{12}\Msolar$}
\item{Ages: 50 Myr $<$ age $<$ 500 Myr}
\item{Extinction: $0.0<A_{V}<1.0$}
\end{itemize}
\noindent
where the adopted parameter ranges are based on the results of the SED
fits presented in Table 2, and are also consistent (in terms of ages) with recent 
results suggesting that reionisation occurred in the redshift interval
$6<z<10$ (Fan et al. 2006; Page et al. 2006). The simulated LBG
catalogue was then subjected to
the same optical and near-infrared selection criteria adopted for the
real data-set (see Section 3). In Fig 5 we show the resulting estimate
for sample completeness as a function of stellar mass. It can be seen
from this figure that for LBGs with stellar masses of 
$\simeq 10^{11}\Msolar$ we are only $\simeq40\%$ complete. Consequently, to calculate our final
estimate of the number density of M $\gtsim\,10^{11}\Msolar$ LBGs within
our sample we are required to correct the observed numbers by a factor of $\simeq2.5$.
Taking this correction into account, our final estimate for the
number density of M~$\gtsim\,10^{11}\Msolar$ LBGs at $z\geq5$ is 
$\log(\phi/{\rm Mpc}^{-3}) = -5.2^{+0.2}_{-0.3}$, where the quoted
error includes a 20\% contribution from cosmic variance (Somerville et
al. 2004), but does not include possible systematics in the adopted
completeness correction. If, in reality, LBG candidate MCD3 is an M dwarf
contaminant, then the estimate number density would be $\simeq20$\% lower.

\begin{figure} 
\centerline{\epsfig{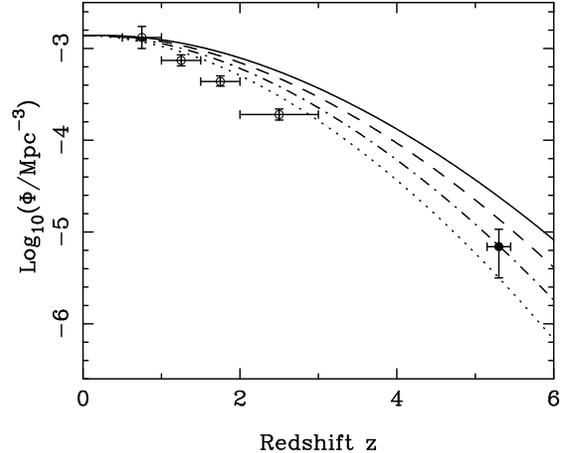}}
\caption{The filled circle at $z=5.3$ is our estimate for
the number density of galaxies with stellar masses greater than
$\simeq 10^{11}\Msolar$ from this study. The data-points at $z\leq3$ (open circles) are the estimated
number densities of galaxies with stellar masses M $\geq
10^{11}\Msolar$ from the GOODS CDFS $K-$band selected sample of Caputi
et al. (2006). The curves show the redshift evolution of the number density
of dark matter halos with masses M~$\geq
1.5\times10^{12}\Msolar$. The solid, dashed, dot-dash and dotted
curves correspond to four different values of $\sigma_{8}$
(0.90, 0.85, 0.80 \& 0.75 respectively).}
\end{figure}

\subsection{Comparison with dark matter halo number densities}
The most fundamental test which we can perform is to ask whether 
$\Lambda$CDM produces the required number of dark matter halos to host
galaxies with stellar masses of $\simeq10^{11}\Msolar$ at this early epoch. In order to
explore this issue we have calculated the number density of
dark matter halos as a function of redshift predicted by the 
Press-Schechter formalism (Bond et al. 1991) \footnote{using code
kindly provided by Will Percival}. In Fig 6 we show the redshift 
evolution of the dark matter halo number density for halos more massive than
$1.5\times10^{12}\Msolar$, which corresponds to a stellar to dark matter ratio of 15. 
Although the typical stellar to dark matter ratio of massive galaxies as
a function of redshift is not accurately known, our choice of 15 as a
representative figure is motivated by several lines of reasoning. Firstly,
assuming a halo occupation number of unity, the predicted number
density of M$_{\rm DM}\geq1.5\times10^{12}\Msolar$ dark matter halos at $z=0$
is in good agreement with the locally observed number density of
galaxies with stellar masses $\geq 10^{11}\Msolar$ (Salpeter IMF; Cole
et~al. 2001). Moreover, a ratio between halo mass and stellar mass of
$\simeq15$ at low-redshift is in good agreement with recent
galaxy-galaxy lensing results (Mandelbaum et al. 2006). Finally, at
high redshift, a stellar to dark matter ratio of $\simeq15$ is also in
agreement with the predictions of recent galaxy formation models
(R. Bower, private communication). At high redshift the number density of dark matter halos
is strongly dependent on the value of the $\sigma_{8}$
parameter. Consequently, in Fig 6 we show the dark matter halo number
density evolution for four different values of $\sigma_{8}$ (0.90,
0.85, 0.80 \& 0.75). The final value of $\sigma_{8}=0.75$ is in
agreement with the WMAP three year results (Spergel et al. 2006).

In Fig 6 we also plot our estimated number
density of LBGs with stellar masses $\gtsim\,10^{11}\Msolar$ from this 
study (median $z=5.3$). It can be seen from Fig 6 that, assuming a
stellar to dark matter ratio of 15, $\Lambda$CDM does produce sufficient
halos at high redshift to host our LBG candidates provided $\sigma_{8}\geq0.8$. In fact, given the
large uncertainties associated with our estimated number density, our
results are also fully consistent with the latest determination of
$\sigma_{8}=0.75$. Consequently, we can conclude that our estimated
number density of massive galaxies at $z\geq5$ can certainly be
accommodated within current $\Lambda$CDM models of structure formation. 

Furthermore, it has recently been suggested that $\simeq60\%$ of the stellar mass at
$z\simeq5$ is missed by traditional drop-out selection
techniques (Stark et al. 2006), which exclude high-redshift galaxies
which are too red in the rest-frame optical to satisfy LBG selection 
criteria (see discussion in Section 6). However, even if we increase
our estimated number density of massive galaxies at $z\simeq5$ by a
factor of two, and assume $\sigma_{8}=0.75$, $\Lambda$CDM can still
produce the required number of dark matter halos if the stellar to dark
matter ratio is $\simeq 10$.

In Fig 6 we also show the estimated number densities of
$\geq10^{11}\Msolar$ galaxies at $z\leq3$ derived from a $K-$band 
selected sample of galaxies in the GOODS CDFS (Caputi et al. 2006). 
The Caputi et al. sample has a magnitude limit of $K\leq23.3$, and is 
$\geq85$\% complete for stellar masses
$\geq10^{11}\Msolar$ (Salpeter IMF) at redshifts $z\leq3$. As can be seen from
Fig 6, the conclusions of Caputi et al. were that $\simeq20$\% of the
local mass density comprised by galaxies with stellar masses
$\geq10^{11}\Msolar$ was already in place by $z\simeq2$, and that the
majority was in place by $z\simeq1$. Similar
conclusions have been reached by numerous studies of deep, small area,
near-infrared surveys (e.g. Drory et al. 2005; Fontana 2004). The
results presented here strongly suggest that the increase 
in number density of massive galaxies within the redshift interval 
$2.5<z<5.5$ closely traces the build-up of suitable dark matter halos
and that, as a consequence, only $\ltsim\,1\%$ of the local density of
$\geq10^{11}\Msolar$ galaxies was in place by $z\simeq5$.

\begin{figure}
\centerline{\epsfig{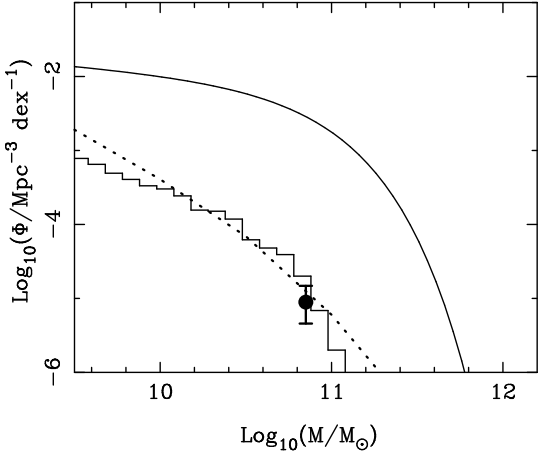}}
\caption{The filled circle is our estimate of the number
density of massive galaxies at $z=5.3$ (see text for discussion). The lower solid line is
the number density of galaxies as a function of stellar mass at
$z=5.3$ from the latest version of the Durham semi-analytic 
galaxy formation model, which incorporates AGN feedback (Bower et al. 2006). 
The dotted curve is the latest prediction of the galaxy formation
model described in Granato et al. (2004). The upper smooth curve shows the number density
of galaxies as a function of stellar mass at $z\simeq0$ (Kennicutt IMF; Cole et al. 2001).}
\end{figure}

\subsection{Comparison with galaxy formation models}
It has recently become apparent that the number density of massive,
red galaxies observed at high redshift is much larger than would have
been predicted by semi-analytic galaxy formation models five years
ago (e.g. Cimatti et al. 2002; Glazebrook et al. 2004). In essence this
is simply a result of the fact that galaxy formation models based on 
hierarchical assembly naturally produce the most massive galaxies at
late times. This apparent discrepancy between the models and
observations has been described variously as ``anti-hierarchical''
behaviour, or cosmic downsizing. It is clear therefore that important
constraints can be placed on the latest generation of galaxy formation
models via comparison with the observed number density of massive
galaxies already in place at $z\geq5$.

In Fig 7 we show the predicted number density of galaxies as a
function of stellar mass at $z=5.3$ (median redshift of our LBG
candidates) from the latest generation of Durham semi-analytic galaxy 
formation models, which incorporate AGN feedback (Bower et al. 2006). 
Also shown in Fig 7 are the predictions (G.L. Granato, private
communication) from the latest version of the Granato et al. (2004)
galaxy formation model, which is based on an anti-hierarchical baryon 
collapse scenario. The filled circle is our estimate for the number
density of galaxies within a bin centred on $10^{11}\Msolar$, based on the results of this
study. When plotting the data-point in Fig 7, we have shifted the
centre of the bin to lower mass by 0.15 dex, to account for the
difference between the Salpeter IMF adopted in this paper, and the
IMFs adopted in the Bower et al. and Granato et al. models (Kennicutt
and Romano et al. 2002 respectively). As can be seen from Fig 7, our
estimated number density is clearly in
good agreement with the predictions of the Bower et al. and 
Granato et al. models. Interestingly, given that our number
density is almost certainly a lower limit, both model predictions can
accommodate a number density which is a factor of $\simeq2-3$ higher than our
estimate, as required if LBGs constitute $\leq50\%$ of the total
stellar mass at $z\geq5$ (e.g. Stark et al. 2006). Finally, although
not included in Fig 7, we also note that our
estimated number density of $z^{\prime}\leq25$ LBGs at $z\geq5$ is 
fully consistent with the hydro-dynamical simulations of Night et al. (2006).

\section{Comparison with recent results}
In this study we have exploited the large co-moving volume covered by
the SXDS/UDS data-set to identify a sample of bright
($z^{\prime}\leq25$) LBGs at $z\geq5$. In this section we compare our
results with those of three recent studies based on the extensive
multi-wavelength data available in the GOODS CDFS (Grazian et
al. 2006; Stark et al. 2006; Yan et al. 2006).

The GOODS-MUSIC sample (Grazian et al. 2006) consists of $z-$band and
$K-$band selected samples in the southern GOODS field, and provides
a photometric redshift for each object calculated from SED fits to the extensive,
high-quality, HST+VLT+Spitzer imaging available in the field. 
The $z-$band catalogue is $100\%$ complete to $z_{850}=25$ and,
although it only covers 142 square arcmin, it is obviously of some
interest to compare the number of $z\geq5$ galaxies found in the MUSIC
sample to the results presented here. Based on the results of this
study, and correcting for the difference
in area, the number of $z_{850}\leq25$ LBG-type galaxies within the
GOODS-MUSIC catalogue should be consistent with $0.73\pm0.30$.
Interestingly, the GOODS-MUSIC catalogue
contains seven objects with $z_{850}\leq25$ and a photometric redshift
estimate of $z\geq5$, seemingly very inconsistent
with the results presented here. However, our own SED fits to the
MUSIC photometry for these seven objects suggest that only two are
believable high-redshift candidates.

Three of the seven objects (MUSIC
IDs 1133, 8316 \& 12966) have SEDs which are clearly stellar in nature, and
have SExtractor stellarity parameters of 0.98, 0.99 and 0.99
respectively. In addition, MUSIC ID=7004 has a clear $V-$band
detection which is inconsistent with the best-fitting photometric 
redshift of $z_{phot}=6.91$. Finally, MUSIC ID=10140 has a very
unusual SED which would appear to be due to severe blending with
bright nearby objects in the near-infrared and Spitzer bands. 
Of the original seven objects, this leaves two apparently robust $z\geq5$
candidates with $z_{850}\leq25$ (MUSIC IDs 499 \& 3094). One of these (MUSIC ID=499) is the 
spectroscopically confirmed LBG SBM03\#03 (Bunker et al. 2003) and the
other (MUSIC ID=3094) has been spectroscopically confirmed at
$z=5.55$ by the GOODS VLT spectroscopy campaign (Vanzella et
al. 2006). Consequently, given that
the GOODS MUSIC catalogue is complete to $z_{850}=25$
(i.e does not employ any colour selection), and the much larger cosmic
variance in the smaller GOODS field ($\simeq40\%$; Somerville et
al. 2004), we conclude that the number of robust $z\geq5$ galaxies
with $z_{850}\leq25$ in the MUSIC catalogue is in good agreement with
the results presented here.

In their recent study, Stark et al. (2006) assemble a combined
photometric and spectroscopic catalogue of objects in the GOODS-CDFS
field to investigate the stellar mass density at $z\simeq5$. Within
their sample Stark et al. have two objects with $z\geq5$ and
$z_{850}\leq25$, one with a spectroscopic redshift of $z=5.55$ (MUSIC
ID=3094) and the other with a photometric redshift of $z_{phot}=5.38$
(also from the MUSIC catalogue). Our SED fits to these two objects successfully
reproduce the redshift of the spectroscopically confirmed candidate, but are
unable to produce a statistically acceptable high-redshift fit to the
candidate with $z_{phot}=5.38$. Consequently, if we also include SBM03\#03
($z_{spec}=5.78$; Bunker et al. 2003), we again conclude that the
GOODS-CDFS field contains two $z\geq5$ objects with $z_{850}\leq25$.

Interestingly, the $z_{spec}=5.55$ object in the Stark et
al. sample is not a classic $i-$drop ($i_{814}-z_{850}=0.57$)
and would not have passed our strict selection criteria (estimated
colour $R-z^{\prime}=2.1$ in the Subaru filters). This has two
consequences. Firstly, it brings the observed number of $z\geq5$ LBGs
in the GOODS-CDFS field with $z_{850}\leq 25$ into good agreement
with the results presented here, which correspond to $0.73\pm0.30$
objects per GOODS-CDFS area. Secondly, it provides further evidence for the conclusion
of Stark et al. (2006) that traditional drop-out selection techniques
are missing $\simeq60\%$ of the stellar mass at $z\simeq5$. Finally,
Yan et al. (2006) have recently completed a study of the
stellar mass density at $z\simeq6$ based on $i-$drop selection of
galaxies in the north and south GOODS fields (combined area $\simeq
330$ square arcmin). Within their sample, Yan et al. find two objects
at $z\geq5$ with $z_{850}\leq25$. This is in good agreement with the results
presented in this paper, which correspond to $1.69 \pm 0.56$
objects per GOODS field (N+S).

\subsection{The absence of super-massive galaxies}
As a final comment we note here that we have found no evidence in this
study for the existence of LBGs with masses in excess of
$3\times10^{11}\Msolar$ in the redshift interval $5<z<6$. Although
still uncertain, this non-detection is notable given that we should
be $\simeq60\%$ complete to LBG-type objects of this mass (see Fig 5) over a
volume of $\simeq 5\times10^{6}$Mpc$^{3}$. Consequently, the detection
of even one object of this mass within the UDS/SXDS field would have
pointed to a number density of $\log(\phi/{\rm Mpc}^{-3}) \simeq
-6.7$. Our non-detection is in agreement with the recent search for 
$\geq3\times10^{11}\Msolar$ galaxies within the GOODS-CDFS field by
Dunlop, Cirasuolo \& McLure (2006).

\section{Conclusions}
In this paper we have presented the results of a study aimed at
exploiting the large co-moving volume of the SXDS/UDS data-set to
identify the most massive LBGs at $z\geq5$. Our main conclusions can be
summarised as follows:

\begin{enumerate}

\item{We have identified a robust sample of nine bright
($z^{\prime}\leq~25$) LBG candidates with photometric redshifts in the
range $5.1<~z~<5.9$.} 

\item{Our corresponding estimate of the surface density for
$z^{\prime}\leq~25$ LBGs at $z\geq5$ is $0.005\pm0.002$ per
square~arcmin. This surface density estimate is in good agreement with,
although much more robust than, a prediction based on the one
previously known LBG candidate with $z_{850}\leq25$ in a compilation of the
deep HST fields; $0.004\pm0.004$ per square arcmin (Bouwens et al. 2006).}

\item{SED fits to the stacked photometry of the final nine LBG
candidates suggests that their typical stellar mass is
$\gtsim\,5\times10^{10}\Msolar$ (Salpeter IMF), comparable with the
most massive galaxies currently known at $z\geq5$.}

\item{Based on our best-fitting stellar mass estimates, five of the
nine LBG candidates have estimated stellar masses
$\gtsim\,10^{11}\Msolar$. Our corresponding estimate for the number
density of $z\geq5$ LBGs with stellar masses $\gtsim\,10^{11}\Msolar$ is 
$\log(\phi/{\rm Mpc}^{-3}) = -5.2^{+0.2}_{-0.3}$.}

\item{This number density is found to be compatible with the 
density of suitable dark matter halos predicted by $\Lambda$CDM
structure formation models for plausible combinations of $\sigma_{8}$
and the ratio of stellar to dark matter.}

\item{It is found that recent galaxy formation models can account for 
the existence of such massive galaxies at high redshift, even if LBGs only account for
$\leq50\%$ of the total stellar mass at $z\geq5$.}

\item{Despite the large co-moving volume surveyed, no evidence is
found for the existence of LBGs with masses in 
excess of $3\times~10^{11}\Msolar$ at $z\geq5$.}

\end{enumerate}

\section{acknowledgements}
The authors would like to acknowledge Richard Bower and Gian Luigi
Granato for providing their stellar mass function predictions. The
authors would also like to acknowledge Nigel Hambly and Niall Deacon
for useful discussions on the nature of cool galactic stars and Will
Percival for providing the code to calculate halo number densities. 
RJM, OA and IS would like to acknowledge the funding of the Royal
Society. MC, SF and CS would like to acknowledge funding from
PPARC. We are grateful to the staff at UKIRT and Subaru for making these 
observations possible. We also acknowledge the Cambridge Astronomical
Survey Unit and the Wide Field Astronomy Unit in Edinburgh for processing the UKIDSS data.

\end{document}